\documentclass[nofootinbib,preprint,preprintnumbers,a4paper,11pt]{revtex4}
\usepackage{epsfig}
\usepackage{footnote}
\usepackage{ulem}
\usepackage{color}
\usepackage{array}
\usepackage{amssymb}
\usepackage{amsmath}
\usepackage{graphicx}
\usepackage{subfigure}
\usepackage{longtable}
\usepackage{verbatim}
\usepackage{amsfonts}
\usepackage{hyperref}
\usepackage{multirow}
\usepackage{cancel}

\begin{document}

\title{Generation of isospin sum rules in heavy hadron weak decays}

\author{Di Wang$^{1}$}\email{wangdi@hunnu.edu.cn}

\address{%
$^1$Department of Physics, Hunan Normal University, and Key Laboratory of Low-Dimensional Quantum Structures and Quantum Control of Ministry of Education, Changsha 410081, China
}

\begin{abstract}
Isospin symmetry is the most precise flavor symmetry.
In this work, we propose an approach to generate isospin sum rules for heavy hadron decays without the Wigner-Eckhart invariants.
The effective Hamiltonian of heavy quark weak decay is fully invariant under a series of isospin lowering operators $I_-^n$ and then the isospin sum rules can be generated through several master formulas.
It provides a systematic way to study the isospin symmetry of $c$- and $b$-hadron weak decays.
The theoretical framework of this approach is presented in detail with the nonleptonic decays of $D$ and $B$ mesons as examples.
In addition, the $V$-/$U$-spin sum rules are derived in a similar algorithm by replacing $I_-^n$ with $V_-^n$/$U_-^n$.
\end{abstract}

\maketitle


\section{Introduction}

Flavor symmetry, as a powerful tool to analyze heavy meson and baryon weak decays, has been extensively studied in literature~\cite{Qin:2021zqx,Qin:2020zlg,Gronau:2000zy,Chau:1993ec,Muller:2015lua,Buccella:2019kpn,Feldmann:2012js,
Cheng:2012xb,Jung:2009pb,Grossman:2003qp,Ligeti:2015yma,
Franco:2012ck,Bhattacharya:2012ah,Brod:2012ud,Muller:2015rna,Grossman:2013lya,
Hiller:2012xm,Pirtskhalava:2011va,Grossman:2006jg,Falk:2001hx,Chau:1991gx,
Savage:1991wu,Voloshin:1975yx,Quigg:1979ic,Golden:1989qx,Abbott:1979fw,
Altarelli:1974sc,Einhorn:1975fw,Kingsley:1975fe,Wang:2022fbk,Wang:2022yyn,
Qin:2022nof,He:2018joe,He:2018php,Wang:2020gmn,Wang:2022wrb,Hsiao:2020iwc,
Zhao:2018mov,Groote:2021pxt,Han:2021azw,
Chau:1995gk,Kohara:1991ug,Geng:2019awr,He:2021qnc,Li:2021rfj,Geng:2020zgr,
Geng:2018rse,Geng:2019bfz,Geng:2019xbo,
Hsiao:2019yur,Jia:2019zxi,Wang:2019dls,Savage:1989qr,Sheikholeslami:1991ab,
Sharma:1996sc,Verma:1995dk,Wang:2017azm,Shi:2017dto,Wang:2018utj,Lu:2016ogy,
Geng:2017esc,Geng:2017mxn,Wang:2017gxe,Geng:2018plk,Geng:2018bow,Gavrilova:2022hbx,
Hassan:2022ucn}.
It leads to linear relations between amplitudes of some hadronic processes, known as flavor sum rules.
Isospin symmetry is the most precise flavor symmetry.
Isospin breaking is naively expected as $\delta_I \simeq (m_u-m_d)/\Lambda_{\rm QCD}\sim 1\%$, while $V$/$U$-spin breaking is $\delta_{V/U} \simeq m_s/\Lambda_{\rm QCD}\sim 30\%$.
Isospin sum rules could provide knowledge on unmeasured channels and be used to  extract useful information of hadronic dynamics.
For instance, the isospin sum rule of $B\to\pi\pi$ system is critical in the Gronau-London method \cite{Gronau:1990ka} of determining the Cabibbo-Kobayashi-Maskawa (CKM) quark mixing parameter $\alpha\equiv Arg[V_{td}V^*_{tb}/V_{ud}V^*_{ub}]$.

Flavor sum rules are usually found by observing decay amplitudes expressed by the Wigner-Eckhart invariants \cite{Eckart30,Wigner59}.
For example, the isospin sum rule of $\overline B\to \pi\pi$ system is derived from the isospin decompositions of $B^-\to \pi^0\pi^-$, $\overline B^0\to \pi^+\pi^-$ and $\overline B^0\to \pi^0\pi^0$ modes.
A useful method for generating $SU(3)$ sum rules for charm meson decays without the Wigner-Eckhart invariants was proposed in \cite{Grossman:2012ry}.
The effective Hamiltonian of charm quark decay is invariant under operators $T_-$ and $S$, which allows us to generate $SU(3)$ sum rules through several master formulas.
This approach has been extended to singly and doubly charmed baryon decays~\cite{Wang:2022kwe}.
However, $T_-$ is a linear combination of isospin and $V$-spin operators and $S$ is a linear combination of three $U$-spin operators. Isospin sum rules cannot be generated by $T_-$ and $S$.

In this work, we propose an approach to generate isospin sum rules by a series of isospin lowering operators $I_-^n$. Isospin sum rules are derived though several master formulas.
Taking the nonleptonic decays of $D$ and $B$ mesons as examples, our method is shown in detail.
The $V$- and $U$-spin sum rules can also be derived in a similar algorithm by replacing $I_-^n$ with $V_-^n$ and $U_-^n$.
This approach could be easily applied to other decay modes such as heavy baryon decays, multi-body decays, etc. It provides a systematic way to analyze flavor symmetry in $c$- and $b$-hadron decays.

The rest of this paper is structured as follows.
In Sec. \ref{ID}, the $D\to PP$ modes are selected as examples to introduce the theoretical framework of generating isospin sum rules.
The isospin sum rules of $B$ meson decays are discussed in Sec. \ref{IB}.
Sec. \ref{summary} is a short summary.
And the $V$- and $U$-spin sum rules are investigated in Appendices \ref{Vsum} and \ref{Usum}, respectively.

\section{Isospin sum rules in the $D\to PP$ decays}\label{ID}
In this section, we present our theoretical framework for generating isospin sum rules, taking the nonleptonic $D$ meson decays as examples.
The decays of charm quark are classified as Cabibbo-favored (CF), singly Cabibbo-suppressed (SCS), and doubly Cabibbo-suppressed (DCS) decays.  The flavor structures of CF, SCS and DCS decays are $c\to s\bar d u$, $c\to  d\bar du/ s \bar su$, $c\to d\bar s u$, respectively.
For CF decay, isospin and its third component change as $\Delta I = 1$, $\Delta I_3 = 1$.
For SCS decay, isospin and its third component change as $\Delta I = 3/2$ or $1/2$, $\Delta I_3 = 1/2$.
And for DCS decay, isospin and its third component change as $\Delta I = 1$ or $0$, $\Delta I_3 = 0$.
There exists a basis in which isospin sum rules involve only CF, SCS or DCS decays respectively.

The effective Hamiltonian of charm quark decay in the Standard Model (SM) is \cite{Buchalla:1995vs}
\begin{align}\label{hsm}
 \mathcal H_{\rm eff}={\frac{G_F}{\sqrt 2} }
 \left[\sum_{q=d,s}V_{cq_1}^*V_{uq_2}\left(\sum_{i=1}^2C_i(\mu)O_i(\mu)\right)
 -V_{cb}^*V_{ub}\left(\sum_{i=3}^6C_i(\mu)O_i(\mu)+C_{8g}(\mu)O_{8g}(\mu)\right)\right]+h.c.,
 \end{align}
where the tree operators are
\begin{eqnarray}\label{ot}
O_1=(\bar{u}_{\alpha}q_{2\beta})_{V-A}
(\bar{q}_{1\beta}c_{\alpha})_{V-A},\qquad
O_2=(\bar{u}_{\alpha}q_{2\alpha})_{V-A}
(\bar{q}_{1\beta}c_{\beta})_{V-A},
\end{eqnarray}
with $\alpha,\beta$ being color indices.
The QCD penguin operators are
 \begin{align}\label{op}
 O_3&=(\bar u_\alpha c_\alpha)_{V-A}\sum_{q'=u,d,s}(\bar q'_\beta
 q'_\beta)_{V-A},~~~
 O_4=(\bar u_\alpha c_\beta)_{V-A}\sum_{q'=u,d,s}(\bar q'_\beta q'_\alpha)_{V-A},
 \nonumber\\
 O_5&=(\bar u_\alpha c_\alpha)_{V-A}\sum_{q'=u,d,s}(\bar q'_\beta
 q'_\beta)_{V+A},~~~
 O_6=(\bar u_\alpha c_\beta)_{V-A}\sum_{q'=u,d,s}(\bar q'_\beta
 q'_\alpha)_{V+A}.
 \end{align}
The chromomagnetic penguin operator is
\begin{align}
O_{8g}=\frac{g_s}{8\pi^2}m_c{\bar
u}_\alpha\sigma_{\mu\nu}(1+\gamma_5)T^a_{\alpha\beta}G^{a\mu\nu}c_{\beta},
\end{align}
which can be included into the penguin operators
\cite{Beneke:2003zv,Beneke:2000ry,Beneke:1999br}.
In the $SU(3)$ picture, the effective Hamiltonian of charm quark decay is written as \cite{Wang:2020gmn}
\begin{align}\label{h}
  \mathcal{H}_{\rm eff}= \sum_{i,j,k=1}^{3}H_k^{ij}O^{ij}_k=\sum_{i,j,k=1}^{3}H_k^{ij}(\bar{q}^iq_k)(\bar{q}^jc).
\end{align}
The coefficient matrix $H$ is obtained from the map $(\bar uq_1)(\bar q_2c)\rightarrow V^*_{cq_2}V_{uq_1}$ for current-current operators and $(\bar q^\prime q^\prime)(\bar uc)\rightarrow -V^*_{cb}V_{ub}$ for penguin operators.
Since $q_1$ and $q_2$ could be $d$ or $s$ quark and $q^\prime$ could be $u$, $d$ or $s$ quark according to Eq.~\eqref{hsm}, the non-zero $H^{ij}_k$ induced by tree and penguin operators include
\begin{align}\label{ckm1}
 &\{H^{(0)}\}^{13}_2 = V_{cs}^*V_{ud},  \qquad \{H^{(0)}\}_{2}^{12}=V_{cd}^*V_{ud},\qquad \{H^{(0)}\}_{3}^{13}= V_{cs}^*V_{us}, \qquad \{H^{(0)}\}_{3}^{12}=V_{cd}^*V_{us},\nonumber\\
 &\{H^{(1)}\}^{11}_1 = -V_{cb}^*V_{ub}, \qquad \{H^{(1)}\}^{21}_2=-V_{cb}^*V_{ub}, \qquad \{H^{(1)}\}^{31}_3=-V_{cb}^*V_{ub},
\end{align}
where superscripts $(0)$ and $(1)$ are used to differentiate the tree and penguin contributions.

The light pseudoscalar meson state is expressed as $|P_\alpha\rangle = (P_\alpha)^{i}_{j}|P^{i}_{j} \rangle$, in which $|P^{i}_{j} \rangle$ is the quark composition $|P^{i}_{j} \rangle = |q_i\bar q_j\rangle$ and $(P_\alpha)$ is the coefficient matrix.
In the $SU(3)$ picture, the pseudoscalar meson octet $|P_8 \rangle$ is expressed as
\begin{eqnarray}\label{a1}
 |P_8\rangle =  \left( \begin{array}{ccc}
   \frac{1}{\sqrt 2} |\pi^0\rangle +  \frac{1}{\sqrt 6} |\eta_8\rangle,    & |\pi^+\rangle,  & |K^+\rangle \\
   | \pi^-\rangle, &   - \frac{1}{\sqrt 2} |\pi^0\rangle+ \frac{1}{\sqrt 6} |\eta_8\rangle,   & |K^0\rangle \\
   | K^- \rangle,& |\overline K^0\rangle, & -\sqrt{2/3}|\eta_8\rangle \\
  \end{array}\right).
\end{eqnarray}
The charmed meson state is expressed as $|D_\alpha\rangle = ( |D^0\rangle,\,\, |D^+\rangle ,\,\, |D^+_s\rangle )$.
The decay amplitude of $D_\gamma\to P_\alpha P_\beta$ mode can be constructed as
\begin{align}\label{amp}
\mathcal{A}(D_\gamma\to P_\alpha P_\beta)& =  \langle P_\alpha P_\beta |\mathcal{H}_{\rm eff}| D_\gamma\rangle \nonumber\\& ~= \sum_{\omega}\,(P_\alpha)^n_m\langle P^n_m|(P_\beta)^s_r\langle P^s_r||H^{jk}_lO^{jk}_l||(D_\gamma)_i|D_i\rangle\nonumber\\& ~~=\sum_\omega\,\langle P_{m}^{n} P^s_r |O^{jk}_{l} |D_{i}\rangle \times (P_\alpha)_{m}^{n}(P_\beta)_{r}^{s} H^{jk}_{l}(D_\gamma)_i\nonumber\\& ~~~= \sum_\omega X_{\omega}(C_\omega)_{\alpha\beta\gamma}.
\end{align}
According to the Wigner-Eckhart theorem \cite{Eckart30,Wigner59}, $X_\omega = \langle P_{m}^{n} P^s_r |O^{jk}_{l} |D_{i}\rangle$ is the reduced matrix element that is independent of $\alpha$, $\beta$ and $\gamma$.
All information about initial/final states is absorbed into the Clebsch-Gordan (CG) coefficient $(C_\omega)_{\alpha\beta\gamma}=(P_\alpha)_{m}^{n}(P_\beta)_{r}^{s} H^{jk}_{l}(D_\gamma)_i$.

In general, the flavor sum rules are derived by writing decay amplitudes and combining several modes to form a polygon in the complex plane.
However, this method is laborious and unsystematic.
The authors of Ref.~\cite{Grossman:2012ry} proposed an approach to generate flavor sum rules for charmed meson decays without the Wigner-Eckhart invariants.
The idea is that if there is an operator $T$ under which $TH=0$, it follows that
\begin{equation}\label{rule}
  \langle P_\alpha P_\beta |T\mathcal{H}_{\rm eff}| D_\gamma\rangle = \sum_\omega\,\langle P_{m}^{n} P^s_r |O^{jk}_{l} |D_{i}\rangle \times (P_\alpha)_{m}^{n}(P_\beta)_{r}^{s} (TH)^{jk}_{l}(D_\gamma)_i = 0.
\end{equation}
Operator $T$ can be applied to the initial/final states rather than the effective Hamiltonian.
Then the LHS of Eq.~\eqref{rule} is turned into a sum of several decay amplitudes and Eq.~\eqref{rule} becomes a flavor sum rule.

It is found in Ref.~\cite{Grossman:2012ry} that the effective Hamiltonian is invariant under $T_-$ and $S$, i.e., $T_-H=0$, $SH=0$.
$T_-$ and $S$ are expressed as \cite{Grossman:2012ry}
\begin{eqnarray}
 T_-=  \left( \begin{array}{ccc}
   0   & 0  & 0 \\
     1 &   0  & 0 \\
    \lambda & 0 & 0 \\
  \end{array}\right)\quad {\rm and} \quad S=  \left( \begin{array}{ccc}
   0   & 0  & 0 \\
     0 &   -\lambda  & 1 \\
    0 & -\lambda^2 & \lambda \\
  \end{array}\right),
\end{eqnarray}
where $\lambda$ is a Wolfenstein parameter given by $\lambda\approx 0.225$ \cite{PDG}.
$S$ is a linear combination of $U$-spin operators and $T_-$ is a linear combination of isospin and $V$-spin operators,
\begin{align}
  S = -\lambda U_3 -\lambda^2 U_- + U_+, \qquad T_- = I_- +\lambda V_-.
\end{align}
The $SU(3)$ sum rules generated by $S$ and $T_-$ are $U$-spin sum rules and combinations of isospin and $V$-spin sum rules, respectively.
Besides, the premise of $T_- H=0$ and $S H=0$ is that the CKM matrix elements in charm decay are approximated to be $V_{ud} \approx 1$, $V_{us} \approx \lambda$, $V_{cd} \approx -\lambda$, $V_{cs} \approx 1$.
If the next order correction is included in the Wolfenstein parametrization, we have $T_- H\neq0$ and $S H\neq0$.
So the $SU(3)$ sum rules generated though $S$ and $T_-$ dependent on the Wolfenstein approximation of the CKM matrix elements in charm sector.
$S$ and $T_-$ cannot be used to construct $SU(3)$ sum rules of $b$-hadron decay.

The three operators associated with isospin are $I_3$, $I_+$ and $I_-$.
In this work, we try to establish the master formulas of isospin sum rules through the isospin lowering operator $I_-$, where $I_-$ is expressed as
\begin{eqnarray}
 I_-=  \left( \begin{array}{ccc}
   0   & 0  & 0 \\
     1 &   0  & 0 \\
    0 & 0 & 0 \\
  \end{array}\right).
\end{eqnarray}
To achieve this goal, we decompose the four-quark operator $O^{ij}_k$ to $SU(3)$ irreducible representations, $3 \otimes  \overline 3 \otimes 3 =  3_p\oplus 3_t\oplus  \overline 6 \oplus 15$.
The explicit decomposition is \cite{Grossman:2012ry}
\begin{align}\label{hd}
  O_k^{ij}= \frac{1}{8}\,O(15)_k^{ij}+\frac{1}{4}\,\epsilon^{ijl}O(\overline 6)_{lk}+\delta^j_k\Big(\frac{3}{8}O( 3_t)^i-\frac{1}{8}O(3_p)^i\Big)+
  \delta^i_k\Big(\frac{3}{8}O( 3_p)^j-\frac{1}{8}O( 3_t)^j\Big).
\end{align}
According to the map rule below Eq.~\eqref{h}, the non-zero coefficients corresponding to the tree operators include
\begin{align}\label{ckm3}
 &  \{H^{(0)}( \overline6)\}_{22}=-2\,V_{cs}^*V_{ud},\qquad \{H^{(0)}( \overline 6)\}_{23}=(V_{cd}^*V_{ud}-V_{cs}^*V_{us}),  \qquad \{H^{(0)}( \overline 6)\}_{33}=  2\,V_{cd}^*V_{us},\nonumber \\
   &  \{H^{(0)}(15)\}_{1}^{11}=-2\,(V_{cd}^*V_{ud}+V_{cs}^*V_{us}), \qquad \{H^{(0)}(15)\}_{2}^{13}= 4\,V_{cs}^*V_{ud},  \qquad  \{H^{(0)}(15)\}_{3}^{12}=4\,V_{cd}^*V_{us},\nonumber \\
 &  \{H^{(0)}(15)\}_{2}^{12}= 3\,V_{cd}^*V_{ud}-V_{cs}^*V_{us},\qquad \{H^{(0)}(15)\}_{3}^{13}=3\,V_{cs}^*V_{us}-V_{cd}^*V_{ud}, \nonumber \\ & \{H^{(0)}( 3_t)\}^1=V_{cd}^*V_{ud}+V_{cs}^*V_{us}.
\end{align}
The non-zero coefficients corresponding to the penguin operators include
\begin{align}\label{ckm4}
 \{H^{(1)}( 3_t)\}^1=-V_{cb}^*V_{ub}, \qquad \{H^{(1)}( 3_p)\}^1=-3V_{cb}^*V_{ub}.
\end{align}

The $\overline 6$ representation can be written in matrix form as $[H^{(0)}( \overline6)]^i_j=[H^{(0)}( \overline6)]_{ij}$ with
\begin{eqnarray}
 [H^{(0)}( \overline6)]= \left( \begin{array}{ccc}
   0   & 0  & 0 \\
     0 &   -2\,V_{cs}^*V_{ud}  & (V_{cd}^*V_{ud}-V_{cs}^*V_{us}) \\
    0 & (V_{cd}^*V_{ud}-V_{cs}^*V_{us}) & 2\,V_{cd}^*V_{us} \\
  \end{array}\right).
\end{eqnarray}
Under the isospin lowering operator $I_-$, $[H^{(0)}( \overline6)]$ is transformed as
\begin{align}\label{I6}
 I_-[ H^{(0)}(\overline6)] =I_-\cdot[H^{(0)}( \overline6)]+I_-\cdot[H^{(0)}( \overline 6)]^T = 0,
\end{align}
where symbol " $\cdot$ " represents the dot product of two matrices and superscript $T$ represents the transposition of matrix.
The three $3$-dimensional presentations are written in matrix form as
\begin{align}
 [H^{(0)}( 3_t)]&= ( \,V_{cd}^*V_{ud}+V_{cs}^*V_{us}, \,0,\, 0\, ),\\
[H^{(1)}( 3_t)]&= (\,-V_{cb}^*V_{ub},\, 0,\, 0\,),\\
[H^{(1)}( 3_p)]&= (\,-3V_{cb}^*V_{ub},\, 0,\,0\,).
\end{align}
Under the isospin lowering operator $I_-$, $[H^{(0,1)}(3_{t,p})]$ are transformed as
\begin{align}\label{I3}
 I_-[H^{(0,1)}( 3_{t,p})] =[H^{(0,1)}( 3_{t,p})]\cdot I_- = 0.
\end{align}
One can find $[H^{(0)}( \overline6)]$ and $[H^{(0,1)}(3_{t,p})]$ are zero under $I_-$.
The $\underline{15}$ representation is written in matrix form as $\{[H^{(0)}(15)]_i\}_j^k = [H^{(0)}(15)]_{k}^{ij}$ with
\begin{eqnarray}
 [H^{(0)}(15)]_1= \left( \begin{array}{ccc}
   -2\,(V_{cd}^*V_{ud}+V_{cs}^*V_{us})   & 0  & 0 \\
     0 &   (3\,V_{cd}^*V_{ud}-V_{cs}^*V_{us})  & 4\,V_{cs}^*V_{ud} \\
    0 & 4\,V_{cd}^*V_{us} & (3\,V_{cs}^*V_{us}-V_{cd}^*V_{ud}) \\
  \end{array}\right),
\end{eqnarray}
\begin{eqnarray}
[H^{(0)}(15)]_2= \left( \begin{array}{ccc}
   0   & 0  & 0 \\
     (3\,V_{cd}^*V_{ud}-V_{cs}^*V_{us}) &   0  & 0 \\
    4\,V_{cd}^*V_{us} & 0 & 0 \\
  \end{array}\right),
\end{eqnarray}
\begin{eqnarray}
[H^{(0)}(15)]_3= \left( \begin{array}{ccc}
   0   & 0  & 0 \\
     4\,V_{cs}^*V_{ud} &   0  & 0 \\
    (3\,V_{cs}^*V_{us}-V_{cd}^*V_{ud}) & 0 & 0 \\
  \end{array}\right).
\end{eqnarray}
The tensor transformation law of $[H^{(0)}(15)]$ under $I_-$ is
\begin{align}\label{15law}
 \{I_-[H^{(0)}(15)]_{i}\}^k_j=2\,\{[H^{(0)}(15)]_{(i}\cdot I_-\}^k_{j)}-\{I_-\cdot[H^{(0)}(15)]_{i}\}^k_j.
\end{align}
Under the isospin lowering operator $I_-$, $[H^{(0)}(15)]_{2,3}$ are zero but $[H^{(0)}(15)]_{1}$ is non-zero,
\begin{align}\label{I15x}
 I_-[H^{(0)}(15)]_{2,3} =0,\qquad
 I_-[H^{(0)}(15)]_1 =\left( \begin{array}{ccc}
   0   & 0  & 0 \\
     8\,V_{cd}^*V_{ud} &   0  & 0 \\
    8\,V_{cd}^*V_{us} & 0 & 0 \\
  \end{array}\right) \neq 0.
\end{align}
Thereby, the isospin lowering operator $I_-$ cannot be used to construct flavor sum rules directly like $T_-$ and $S$.

Notice the matrix $I_-[H^{(0)}(15)]_1$ does not include $V^{*}_{cs}V_{ud}$.
By combining with Eqs.~\eqref{I6}, \eqref{I3} and \eqref{I15x}, it is found that the Hamiltonian of CF decay is zero under $I_-$, i.e., $I_-H_{\rm CF}=0$.
Observe that the matrix $I_-[H^{(0)}(15)]_1$ is invariant under $I_-$,
\begin{align}\label{I15z}
I_-^2[H^{(0)}(15)]_1 = I_-\{I_-[H^{(0)}(15)]_1\}=0.
\end{align}
So the Hamiltonian of SCS and DCS decays is zero under $I_-^2$, i.e., $I_-^2\,H_{\rm SCS,DCS}=0$.
In fact, we can define a series of operators $I_-^n$. The Hamiltonian of charm quark decay is zero under $I_-^n$ if $n\geq 2$ since $I_-0=0$. If the operator $T$ in Eq.~\eqref{rule} is replaced by $I_-^n$, Eq.~\eqref{rule} will be an abstract isospin sum rule,
\begin{equation}\label{Isoa}
  \langle P_\alpha P_\beta |I_-^n\mathcal{H}_{\rm eff}| D_\gamma\rangle = \sum_\omega\,\langle P_{m}^{n} P^s_r |O^{jk}_{l} |D_{i}\rangle \times (P_\alpha)_{m}^{n}(P_\beta)_{r}^{s} (I_-^nH)^{jk}_{l}(D_\gamma)_i = 0.
\end{equation}
The derivation of Eqs.~\eqref{I6}, \eqref{I3}, \eqref{I15x} and \eqref{I15z} does not involve the values of CKM matrix elements.
So the isospin sum rules generated from Eq.~\eqref{Isoa} do not rely on any approximation of the CKM matrix.

The abstract isospin sum rule \eqref{Isoa} becomes explicit isospin sum rules by applying $I_-^n$ to initial/final states and computing the coefficients expanded by initial/final states as bases \cite{Wang:2022kwe}.
Under the isospin lowering operator $I_-$, we have
\begin{align}\label{tc3}
&I_-|D_\gamma\rangle = \sum_\alpha |D_\alpha\rangle \langle D_\alpha |I_-|D_\gamma\rangle = \sum_\alpha(D^\alpha)^j [I_-]_j^i(D_\gamma)_i|D_\alpha\rangle = \sum_\alpha\{[I_-]_{D}\}^{\alpha}_{\gamma}|D_\alpha\rangle.
\end{align}
$[I_-]_{D}$ is the coefficient matrix of $I_-|D_\gamma\rangle$ expanded by $|D_\alpha\rangle$, which is derived to be
\begin{eqnarray}
 [I_-]_{D}= \left( \begin{array}{ccc}
   0   & 0  & 0 \\
     1 &  0  & 0 \\
    0 & 0 & 0 \\
  \end{array}\right).
\end{eqnarray}
The isospin lowering operator $I_-$ acting on a pseudoscalar meson octet is a commutator,
\begin{align}\label{m}
I_- \langle [P_8]_\alpha| &= [I_-, \langle [P_8]_\alpha|] =I_-\cdot \langle [P_8]_\alpha | - \langle [P_8]_\alpha | \cdot I_-\nonumber\\&~~= \sum_\beta{\rm Tr}\{[\,I_-,[P_8]_\alpha\,]\cdot [P_8]_\beta^T\}\langle [P_8]_\beta | = \sum_\beta\{[I_-]_{P_8}\}^\beta_\alpha \langle [P_8]_\beta |,
\end{align}
where $[I_-]_{P_8}$ is coefficient matrix of commutator $[\,I_-,\langle[P_8]_\alpha |\,]$ expanded by $\langle [P_8]_\beta |$.
If we define pseudoscalar meson octet as
\begin{align}
 \langle [P_8]_\beta| = ( \langle \pi^+|,\,\,\langle \pi^0|,\,\,\langle \pi^-|,\,\,\langle K^+|,\,\,\langle K^0|,\,\,\langle \overline K^0|,\,\,\langle K^-|,\,\,\langle \eta_8|    ),
\end{align}
we get
\begin{eqnarray}
 [I_-]_{P_8}= \left( \begin{array}{cccccccc}
  0 & 0& 0& 0& 0& 0& 0& 0 \\
  -\sqrt{2}& 0& 0& 0& 0& 0& 0& 0 \\
 0& \sqrt{2}& 0& 0& 0& 0& 0& 0 \\
  0& 0& 0& 0& 0& 0& 0& 0 \\
  0& 0& 0& 1& 0& 0& 0& 0\\
 0& 0& 0& 0& 0& 0& 0& 0\\
 0& 0& 0& 0& 0& -1& 0& 0 \\
 0& 0& 0& 0& 0&0& 0& 0 \\
  \end{array}\right).
\end{eqnarray}
With the matrices $[I_-]_{D}$ and $[I_-]_{P_8}$, the sum of decay amplitudes generated by $I_-$ is written as
\begin{align}\label{rule1}
{ SumI_-}\,[\gamma, \alpha,\beta]= \sum_\mu\left[\{[I_-]_{P_8}\}_\alpha^\mu \mathcal{A}_{ \gamma \to \mu \beta} +  \{[I_-]_{P_8}\}_\beta^\mu \mathcal{A}_{\gamma\to \alpha\mu } + \{[I_-]_{D}\}_\gamma^\mu \mathcal{A}_{\mu\to \alpha \beta }\right].
\end{align}

Since the effective Hamiltonian of CF decay is invariant under $I_-$,
${ SumI_-}\,[\gamma, \alpha,\beta]$ is zero if it is a sum of amplitudes of several CF decay channels.
An isospin sum rule is generated via Eq.~\eqref{rule1} if appropriate $\alpha$, $\beta$ and $\gamma$ are selected.
$I_-$ is the isospin lowering operator. $I_-$ acting on the final/initial state lowers/arises $I_3$ by one.
Then $\alpha$, $\beta$ can be chosen as the states with the maximal $I_3$, and $\gamma$ can be chosen as the state with the minimal $I_3$.
In the $D\to PP$ decays, the choice of $\{\gamma, \alpha, \beta\} = \{D^0, \pi^+,\overline K^0\}$ generates an isospin sum rule as
\begin{align}\label{r1}
{ SumI_-}\,[D^0, \pi^+,\overline K^0]=-\sqrt{2}\,\mathcal{A}(D^0\to \pi^0\overline K^0)-\mathcal{A}(D^0\to \pi^+K^-)+\mathcal{A}(D^+\to \pi^+\overline K^0)=0.
\end{align}
For the SCS and DCS decays, equation $I_-^2H_{\rm SCS,DCS}=0$ indicates that the isospin sum rules are obtained by acting $I_-$ on the final and initial states twice. Specifically, the isospin sum rule of singly Cabibbo-suppressed $D\to \pi\pi$ system is generated by $I_-^2$ with $\{\gamma, \alpha, \beta\} = \{D^0, \pi^+,\pi^+\}$,
\begin{align}\label{r2}
{ SumI_-^2}\,[D^0, \pi^+,\pi^+]&=-\sqrt{2}\,{ SumI_-}\,[D^0, \pi^+,\pi^0]-\sqrt{2}\,{ SumI_-}\,[D^0, \pi^0,\pi^+]+{ SumI_-}\,[D^+, \pi^+,\pi^+]\nonumber\\&=4\,\big[ \mathcal{A}(D^0\to \pi^0 \pi^0)-\mathcal{A}(D^0\to \pi^+\pi^-)-\sqrt{2}\,\mathcal{A}(D^+\to \pi^+ \pi^0)\big]=0.
\end{align}
The isospin rum rule of doubly Cabibbo-suppressed $D\to K\pi$ decays is generated by $I_-^2$ with $\{\gamma, \alpha, \beta\} = \{D^0, \pi^+,K^+\}$,
\begin{align}\label{r3}
{ SumI_-^2}\,[D^0, \pi^+,K^+]&={ SumI_-}\,[D^0, \pi^+,K^0]-\sqrt{2}\,{ SumI_-}\,[D^0, \pi^0,K^+]+{ SumI_-}\,[D^+, \pi^+,K^+]\nonumber\\&~~=-2\,\big[ \sqrt{2}\,\mathcal{A}(D^0\to \pi^0 K^0)+\mathcal{A}(D^0\to \pi^-K^+)\nonumber\\&~~~~~~~~~+\sqrt{2}\,\mathcal{A}(D^+\to \pi^0 K^+)-\mathcal{A}(D^+\to \pi^+ K^0)\big]=0.
\end{align}
The three isospin sum rules derived from $I_-$ and  $I_-^2$ are consistent with the results given by Ref.~\cite{Grossman:2012ry}.

From above analysis, it is found that the isospin sum rules are obtained by applying $I_-^n$ to the initial/final states if the effective Hamiltonian is invariant under $I_-^n$.
Isospin sum rules can also be generated by isospin raising operators $I_+^n$. The results are the same as the ones derived from $I_-^n$.
One should note that not arbitrary choices of $\{\gamma, \alpha, \beta\} $ and $I_-^n$ generate isospin sum rules.
There are two requirements for $\{\gamma, \alpha, \beta\} $ and $I_-^n$. Firstly, the choices of $\{\gamma, \alpha, \beta\} $ and $I_-^n$ should be associated with physical amplitudes.
For example, the choice of $\{\gamma, \alpha, \beta\}  = \{D^+, \pi^+,\overline K^0\}$ and $I_-$ cannot generate an isospin sum rule.
It because that $I_-$ is a QED charge lowering operator and then $I_-$ acting on $\{D^+, \pi^+,\overline K^0\}$ cannot derive charge preserving decay amplitudes.
The choice of $\{\gamma, \alpha, \beta\}  = \{D^0, K^+,K^+\}$ and $I_-^2$ cannot generate an isospin sum rule since $I_-$ does not change strangeness and $\Delta S = -2$ amplitudes are forbidden in charm decay.
Secondly, the isospin sum rule is generated by $I_-^n$ only if $n\geq 1$ for  CF decay and $n\geq 2$ for SCS and DCS decays.
For example, the sum of amplitudes derived by the choice of $\{\gamma, \alpha, \beta\}  = \{D^0, K^+,\overline K^0\}$ and $I_-$ is not zero because it is a sum of SCS amplitudes and $I_-H_{\rm SCS}\neq 0$,
\begin{align}
{ SumI_-}\,[D^0, K^+,\overline K^0]=\mathcal{A}(D^0\to K^0\overline K^0)-\mathcal{A}(D^0\to K^+K^-)+\mathcal{A}(D^+\to K^+\overline K^0)\neq 0.
\end{align}

The change of strangeness in CF, SCS, DCS transitions are $\Delta S = -1$, $0$ and $1$, respectively. $I_-^n$ cannot change strangeness, then we can distinguish the three decay modes though $\Delta S$ in $\gamma \to \alpha\beta$. Considering that $I_-$ is a QED charge lowering operator, we conclude the selection rule of $\{\gamma, \alpha, \beta\}$.
The choice of $\{\gamma, \alpha, \beta\}$ corresponding to a $\Delta Q = 1$ and $\Delta S = -1$ amplitude produces an isospin sum rule of CF mode.
The choice of $\{\gamma, \alpha, \beta\}$ corresponding to a $\Delta Q = 2$ and $\Delta S = 0$ amplitude produces an isospin sum rule of SCS mode.
And the choice of $\{\gamma, \alpha, \beta\}$ corresponding to a $\Delta Q = 2$ and $\Delta S = 1$ amplitude produces an isospin sum rule of DCS mode.
For other choices, no sum rule is generated.
In the $D\to PP$ decays, there are only four choices of $\{\gamma, \alpha, \beta\}$ satisfying above selection rule, $\{\gamma, \alpha, \beta\}$ $=$ $\{D^0, \pi^+,\overline K^0\}$, $\{D^0, \pi^+,\pi^+\}$,  $\{D^0, \pi^+,K^+\}$ and $\{D^+_s, \pi^+,\pi^+\}$.
The first three choices generate the isospin sum rules \eqref{r1}$\sim$\eqref{r3} respectively.
And the choice of $\{\gamma, \alpha, \beta\} = \{D^+_s, \pi^+,\pi^+\}$ generates an isospin sum rule as
\begin{align}\label{t1}
{ SumI_-}\,[D^+_s, \pi^+,\pi^+]=-2\sqrt{2}\,\mathcal{A}(D^+_s\to \pi^+\pi^0)=0.
\end{align}

The approach for generating isospin sum rules can be extended to other decay modes such as $B$ meson decays,  heavy baryon decays, multi-body decays, etc.
It provides a programmatic way to derive isospin sum rules for heavy hadron decays.
In the next section, the applications of our method in the $\overline B\to DP$ and $\overline B\to PP$ decays are discussed.
For the isospin sum rules of other heavy hadron decays and the  phenomenological discussions, we will leave them in the future work.
In addition, the $V$- and $U$-spin sum rules are derived by $V_-^n$ and $U_-^n$ operators with $D\to PP$, $\overline B\to DP$ and $\overline B\to PP$ decays as examples in Appendices \ref{Vsum} and \ref{Usum}.

\section{Isospin sum rules in the $B$ meson decays}\label{IB}

\subsection{Isospin sum rules in the $\overline B\to DP$ decays}\label{IBDP}
The effective Hamiltonian of $b\to c\overline u q$ transition is given by \cite{Buchalla:1995vs}
 \begin{align}\label{hsmb}
 \mathcal H_{\rm eff}={\frac{G_F}{\sqrt 2} }
 \sum_{q=d,s}V_{cb}V_{uq}^*\left[C_1(\mu)O_1(\mu)+C_2(\mu)O_2(\mu)\right]+h.c.,
 \end{align}
where the tree operators are
\begin{eqnarray}
O_1=(\bar{q}_{\alpha}u_{\beta})_{V-A}
(\bar{c}_{\beta}b_{\alpha})_{V-A},\qquad
O_2=(\bar{q}_{\alpha}u_{\alpha})_{V-A}
(\bar{c}_{\beta}b_{\beta})_{V-A}.
\end{eqnarray}
In the $SU(3)$ picture, $O^{i}_j$ is decomposed into irreducible representations as $3 \otimes  \overline 3 =  8\oplus 1$.
The non-zero CKM components include
\begin{align}\label{ckm5}
 &\{H^{(0)}(8)\}^{2}_1 = V_{cb}V_{ud}^*,  \qquad \{H^{(0)}(8)\}_{1}^{3}=V_{cb}V_{us}^*.
\end{align}
$ H^{(0)}(8)$ is written in matrix form as
\begin{eqnarray}
 [H^{(0)}( 8)]= \left( \begin{array}{ccc}
   0   & V_{cb}V_{ud}^*  & V_{cb}V_{us}^* \\
     0 & 0 &  0 \\
    0 & 0 & 0 \\
  \end{array}\right).
\end{eqnarray}
Under $I_-^n$, $[H^{(0)}(8)]$ is transformed as
\begin{align}\label{I8x}
 I_-[ H^{(0)}(8)] =I_-\cdot[H^{(0)}(8)]-[H^{(0)}(8)]\cdot I_- =
  \left( \begin{array}{ccc}
   -V_{cb}V_{ud}^*   & 0  & 0 \\
     0 & V_{cb}V_{ud}^* & V_{cb}V_{us}^* \\
    0 & 0 & 0 \\
  \end{array}\right),
\end{align}
\begin{align}\label{I8y}
 I_-^2[ H^{(0)}(8)] = I_-\{I_-[H^{(0)}(8)]\} =
  \left( \begin{array}{ccc}
   0  & 0  & 0 \\
    -2V_{cb}V_{ud}^* & 0 &  0 \\
    0 & 0 & 0 \\
  \end{array}\right),
\end{align}
\begin{align}\label{I8z}
 I_-^3[ H^{(0)}(8)] = I_-\{I_-\{I_-[H^{(0)}(8)]\}\} =0.
\end{align}
The effective Hamiltonian of $b\to c\overline u d$ ($b\to c\overline u s$) transition is zero under $I_-^n$ with $n\geq 3$ ($n\geq 2$).
So the isospin sum rules of $b\to c\overline u d$ ($b\to c\overline u s$) transition can be generated by $I_-^n$ if $n\geq 3$ ($n\geq 2$).

Under the isospin lowering operator $I_-$, $[I_-]_{\overline B} = [I_-]_{D}$ if the $\overline B$ meson anti-triplet is defined as
 $|\overline B_\alpha\rangle = ( |B^-\rangle,\,\, |\overline B^0\rangle ,\,\, |\overline B^0_s\rangle )$.
The sum of decay amplitudes generated from $\overline B_\gamma\to D_\alpha P_\beta$ under $I_-$ is
\begin{align}\label{rule2}
 { SumI_-}\,[\gamma, \alpha,\beta]=  \sum_\mu\left[\{[I_-]^T_{D}\}_\alpha^\mu \mathcal{A}_{ \gamma \to \mu \beta} +  \{[I_-]_{P_8}\}_\beta^\mu \mathcal{A}_{\gamma\to \alpha\mu } + \{[I_-]_{\overline B}\}_\gamma^\mu \mathcal{A}_{\mu\to \alpha \beta }\right].
\end{align}
The transposition of matrix $[I_-]_{D}$ is arisen from the initial-final transformation of $D$ meson anti-triplet.
With Eq.~\eqref{rule2}, isospin sum rules in the $\overline B\to DP$ modes are derived to be
\begin{align}
{ SumI_-^3}\,[B^-, D^+,\pi^+]&=-{ SumI_-^2}\,[B^-, D^0,\pi^+]-\sqrt{2}\,{ SumI_-^2}\,[B^-, D^+,\pi^0]+{ SumI_-^2}\,[\overline B^0, D^+,\pi^+]\nonumber\\&=2\sqrt{2}\,{ SumI_-}\,[B^-, D^0,\pi^0]-2\,{ SumI_-}\,[B^-, D^+,\pi^-]\nonumber\\&~~~~~~~~-2\,{ SumI_-}\,[\overline B^0, D^0,\pi^+]-2\sqrt{2}\,{ SumI_-}\,[\overline B^0, D^+,\pi^0]\nonumber\\&=6\,\big[ \mathcal{A}(B^-\to D^0\pi^-)+\sqrt{2}\,\mathcal{A}(\overline B^0\to D^0\pi^0)-\mathcal{A}(\overline B^0\to D^+\pi^-)\big]=0,
\end{align}
\begin{align}
{ SumI_-^2}\,[B^-, D^+,\overline K^0]&=-{ SumI_-}\,[B^-, D^0,\overline K^0]-{ SumI_-}\,[B^-, D^+,K^-]+{ SumI_-}\,[\overline B^0, D^+,\overline K^0]\nonumber\\&=2\,\big[ \mathcal{A}(B^-\to D^0K^-)-\,\mathcal{A}(\overline B^0\to D^0\overline K^0)-\mathcal{A}(\overline B^0\to D^+K^-)\big]=0,
\end{align}
\begin{align}\label{t2}
{ SumI_-^2}\,[\overline B^0_s, D^+,\pi^+]&=-{ SumI_-}\,[\overline B^0_s, D^0,\pi^+]-\sqrt{2}\,{ SumI_-}\,[\overline B^0_s, D^+,\pi^0]\nonumber\\&=2\,\big[ \sqrt{2}\,\mathcal{A}(\overline B^0_s\to D^0\pi^0)-\,\mathcal{A}(\overline B^0_s\to D^+\pi^-)\big]=0.
\end{align}

\subsection{Isospin sum rules in the $\overline B\to PP$ decays}\label{IBPP}
The effective Hamiltonian of $b\to u\overline u q$ transition is given by \cite{Buchalla:1995vs}
\begin{align}\label{hsmb2}
 \mathcal H_{\rm eff}=&{\frac{G_F}{\sqrt 2} }
 \sum_{q=d,s}\left[V_{ub}^*V_{uq}\left(\sum_{i=1}^2C_i^u(\mu)O_i^u(\mu)\right) + V_{cb}^*V_{cq}\left(\sum_{i=1}^2C_i^c(\mu)O_i^c(\mu)\right)\right]\nonumber\\&
 -{\frac{G_F}{\sqrt 2}}\sum_{q=d,s}\left[V_{tb}V_{tq}^*\left(\sum_{i=3}^{10}C_i(\mu)O_i(\mu)
 +C_{7\gamma}(\mu)O_{7\gamma}(\mu)+C_{8g}(\mu)O_{8g}(\mu)\right)\right]+h.c..
 \end{align}
The tree operators are
\begin{align}
O_1^u &=(\bar{q}_{\alpha}u_{\beta})_{V-A}
(\bar{u}_{\beta}b_{\alpha})_{V-A},\qquad
O_2^u=(\bar{q}_{\alpha}u_{\alpha})_{V-A}
(\bar{u}_{\beta}b_{\beta})_{V-A},\nonumber\\
O_1^c &=(\bar{q}_{\alpha}c_{\beta})_{V-A}
(\bar{c}_{\beta}b_{\alpha})_{V-A},\qquad
O_2^c=(\bar{q}_{\alpha}c_{\alpha})_{V-A}
(\bar{c}_{\beta}b_{\beta})_{V-A}.
\end{align}
The QCD penguin operators are
 \begin{align}
 O_3&=(\bar q_\alpha b_\alpha)_{V-A}\sum_{q'=u,d,s}(\bar q'_\beta
 q'_\beta)_{V-A},~~~
 O_4=(\bar q_\alpha b_\beta)_{V-A}\sum_{q'=u,d,s}(\bar q'_\beta q'_\alpha)_{V-A},
 \nonumber\\
 O_5&=(\bar q_\alpha b_\alpha)_{V-A}\sum_{q'=u,d,s}(\bar q'_\beta
 q'_\beta)_{V+A},~~~
 O_6=(\bar q_\alpha b_\beta)_{V-A}\sum_{q'=u,d,s}(\bar q'_\beta
 q'_\alpha)_{V+A}.
 \end{align}
The QED penguin operators are
 \begin{align}
 O_7&=\frac{3}{2}(\bar q_\alpha b_\alpha)_{V-A}\sum_{q'=u,d,s}e_{q^\prime}(\bar q'_\beta
 q'_\beta)_{V+A},~~~
 O_8=\frac{3}{2}(\bar q_\alpha b_\beta)_{V-A}\sum_{q'=u,d,s}e_{q^\prime}(\bar q'_\beta q'_\alpha)_{V+A},
 \nonumber\\
 O_9&=\frac{3}{2}(\bar q_\alpha b_\alpha)_{V-A}\sum_{q'=u,d,s}e_{q^\prime}(\bar q'_\beta
 q'_\beta)_{V-A},~~~
 O_{10}=\frac{3}{2}(\bar q_\alpha b_\beta)_{V-A}\sum_{q'=u,d,s}e_{q^\prime}(\bar q'_\beta q'_\alpha)_{V-A}.
 \end{align}
The electromagnetic penguin and chromomagnetic penguin operators are
\begin{align}
O_{7\gamma}&=\frac{e}{8\pi^2}m_b{\bar
q}_\alpha\sigma_{\mu\nu}(1+\gamma_5)F^{\mu\nu}b_\alpha,
\nonumber\\
O_{8g}&=\frac{g_s}{8\pi^2}m_b{\bar
q}_\alpha\sigma_{\mu\nu}(1+\gamma_5)T^a_{\alpha\beta}G^{a\mu\nu}b_{\beta}.
\end{align}
In the $SU(3)$ picture, the coefficient matrices induced by $O_{1,2}^u$ are
\begin{eqnarray}
 [H^{(0,u)}(\overline 6)]= \left( \begin{array}{ccc}
   0   & -V_{ub}V_{us}^*  & V_{ub}V_{ud}^* \\
     -V_{ub}V_{us}^* & 0 &  0 \\
    V_{ub}V_{ud}^* & 0 & 0 \\
  \end{array}\right),
\end{eqnarray}
\begin{eqnarray}
 [H^{(0,u)}(15)]_1= \left( \begin{array}{ccc}
   0   & 3V_{ub}V_{ud}^*  & 3V_{ub}V_{us}^* \\
     0 &   0  & 0 \\
    0 & 0 & 0 \\
  \end{array}\right),
\end{eqnarray}
\begin{eqnarray}
[H^{(0,u)}(15)]_2= \left( \begin{array}{ccc}
   3V_{ub}V_{ud}^*   & 0  & 0 \\
   0 &   -2V_{ub}V_{ud}^*  & -V_{ub}V_{us}^* \\
    0 & 0 & -V_{ub}V_{ud}^* \\
  \end{array}\right),
\end{eqnarray}
\begin{eqnarray}
[H^{(0,u)}(15)]_3= \left( \begin{array}{ccc}
   3V_{ub}V_{us}^*   & 0  & 0 \\
   0 &   -V_{ub}V_{us}^*  & 0 \\
   0 & -V_{ub}V_{ud}^* & -2V_{ub}V_{us}^* \\
  \end{array}\right),
\end{eqnarray}
\begin{align}
 [H^{(0,u)}( 3_t)]= ( \,0, \,\,V_{ub}V_{ud}^*,\,\, V_{ub}V_{us}^*\, ).
\end{align}
The coefficient matrix induced by $O_{1,2}^c$ is
\begin{align}
 [H^{(0,c)}(3_t)]= ( \,0, \,\,V_{cb}V_{cd}^*,\,\, V_{cb}V_{cs}^*\, ).
\end{align}
The coefficient matrices induced by penguin operators are
\begin{align}
 [H^{(1)}(3_t)]= ( \,0, \,\,-V_{tb}V_{td}^*,\,\, -V_{tb}V_{ts}^*\, ),\qquad
[H^{(1)}(3_p)]= ( \,0, \,\,-3V_{tb}V_{td}^*,\,\, -3V_{tb}V_{ts}^*\, ).
\end{align}
One can find all the $3$-dimensional presentations have the structure of $[H(3)]= ( 0, \,a,\, b)$.

Under the operators $I_-^n$, $[H^{(0,u)}(\overline 6)]$, $[H^{(0,u)}(15)]_i$, $[H(3)]$ are transformed as
\begin{align}
 I_-^2[H^{(0,u)}(\overline 6)] =  I_-\{I_-[H^{(0,u)}(\overline 6)]\} =I_-   \left( \begin{array}{ccc}
   0   & 0  & 0 \\
     0 & -2V_{ub}V_{us}^* & 2V_{ub}V_{ud}^* \\
    0 & 0 & 0 \\
  \end{array}\right)=0,
\end{align}
\begin{align}
 I_-^3[H^{(0,u)}(15)]_1 & =  I_-\{I_-\{I_-[H^{(0,u)}(15)]_1\}\} =I_- \left\{I_-  \left( \begin{array}{ccc}
   6V_{ub}V_{ud}^*   & 0  & 0 \\
     0 & -5V_{ub}V_{ud}^* & -4V_{ub}V_{us}^* \\
    0 & 0 & -V_{ub}V_{ud}^* \\
  \end{array}\right)\right\} \nonumber\\&~~=I_-  \left( \begin{array}{ccc}
   0  & 0  & 0 \\
     -16V_{ub}V_{ud}^* & 0 & 0 \\
    0 & 0 & 0 \\
  \end{array}\right)= 0,
\end{align}
\begin{align}
 I_-^2[H^{(0,u)}(15)]_2 & =  I_-\{I_-[H^{(0,u)}(15)]_2\} =I_-  \left( \begin{array}{ccc}
   0   & 0  & 0 \\
     -5V_{ub}V_{ud}^* & 0 & 0 \\
    0 & 0 & 0 \\
  \end{array}\right) = 0,
\end{align}
\begin{align}
 I_-^2[H^{(0,u)}(15)]_3 & =  I_-\{I_-[H^{(0,u)}(15)]_3\} =I_-  \left( \begin{array}{ccc}
   0   & 0  & 0 \\
     -4V_{ub}V_{us}^* & 0 & 0 \\
    -V_{ub}V_{ud}^* & 0 & 0 \\
  \end{array}\right) = 0,
\end{align}
\begin{align}
 I_-^2[H(3)] =  I_-\{I_-[H(3)]\} =I_-\,( \begin{array}{ccc}
   a   & 0  & 0
  \end{array}) = 0.
\end{align}
The isospin sum rules of $b\to u\overline u d$ ($b\to u\overline u s$) transition can be generated by $I_-^n$ in the case of $n\geq 3$ ($n\geq 2$).
The sum of decay amplitudes generated from $\overline B_\gamma\to P_\alpha P_\beta$ under $I_-$ is
\begin{align}\label{rule3}
 { SumI_-}\,[\gamma, \alpha,\beta]=  \sum_\mu\left[\{[I_-]_{P_8}\}_\alpha^\mu \mathcal{A}_{ \gamma \to \mu \beta} +  \{[I_-]_{P_8}\}_\beta^\mu \mathcal{A}_{\gamma\to \alpha\mu } + \{[I_-]_{\overline B}\}_\gamma^\mu \mathcal{A}_{\mu\to \alpha \beta }\right].
\end{align}
With Eq.~\eqref{rule3}, the isospin sum rules of $\overline B\to PP$ modes are derived to be
\begin{align}
{ SumI_-^3}\,[B^-, \pi^+,\pi^+]&=-\sqrt{2}\,{ SumI_-^2}\,[B^-, \pi^0,\pi^+]-\sqrt{2}\,{ SumI_-^2}\,[B^-, \pi^+,\pi^0]+{ SumI_-^2}\,[\overline B^0, \pi^+,\pi^+]\nonumber\\&=-2\,{ SumI_-}\,[B^-, \pi^+,\pi^-]+4\,{ SumI_-}\,[B^-, \pi^0,\pi^0]-2\,{ SumI_-}\,[B^-, \pi^-,\pi^+]\nonumber\\&~~~~~~~~-2\sqrt{2}\,{ SumI_-}\,[\overline B^0, \pi^+,\pi^0]-2\sqrt{2}\,{ SumI_-}\,[\overline B^0, \pi^0,\pi^+]\nonumber\\&=12\,\big[\sqrt{2}\,\mathcal{A}(B^-\to \pi^0\pi^-)+\,\mathcal{A}(\overline B^0\to \pi^0\pi^0)-\mathcal{A}(\overline B^0\to \pi^+\pi^-)\big]=0,
\end{align}
\begin{align}
{ SumI_-^2}\,[B^-, \pi^+,\overline K^0]&=-\sqrt{2}\,{ SumI_-}\,[B^-, \pi^0,\overline K^0]-{ SumI_-}\,[B^-, \pi^+,K^-]+{ SumI_-}\,[\overline B^0, \pi^+,\overline K^0]\nonumber\\&=2\,\big[ \sqrt{2}\,\mathcal{A}(B^-\to \pi^0K^-)-\mathcal{A}(B^-\to \pi^-\overline K^0)\nonumber\\&~~~~~-\sqrt{2}\,\mathcal{A}(\overline B^0\to \pi^0\overline K^0)-\mathcal{A}(\overline B^0\to \pi^+K^-)\big]=0,
\end{align}
\begin{align}\label{t3}
{ SumI_-^2}\,[\overline B^0_s, \pi^+,\pi^+]&=-\sqrt{2}\,{ SumI_-}\,[\overline B^0_s, \pi^0,\pi^+]-\sqrt{2}\,{ SumI_-}\,[\overline B^0_s, \pi^+,\pi^0]\nonumber\\&=4\,\big[\, \mathcal{A}(\overline B^0_s\to \pi^0\pi^0)-\,\mathcal{A}(\overline B^0_s\to \pi^+\pi^-)\big]=0.
\end{align}

According to Eqs.~\eqref{t1}, \eqref{t2} and \eqref{t3}, the branching fractions of $D^+_s\to \pi^+\pi^0$, $\overline B^0_s\to D^0\pi^0$, $\overline B^0_s\to D^+\pi^-$, $\overline B^0_s\to \pi^+\pi^-$ and $\overline B^0_s\to \pi^0\pi^0$ satisfy following equations under isospin symmetry,
\begin{align}
 \mathcal{B}r(D^+_s\to \pi^+\pi^0)&=0,\\ \mathcal{B}r(\overline B^0_s\to D^+\pi^-)&=2\,\mathcal{B}r(\overline B^0_s\to D^0\pi^0),\\
 \mathcal{B}r(\overline B^0_s\to \pi^+\pi^-)&=2\,\mathcal{B}r(\overline B^0_s\to \pi^0\pi^0),
\end{align}
where the identical factor in the $\overline B^0_s\to \pi^0\pi^0$ channel is considered.
So we suggest to measure the branching fractions of $D^+_s\to \pi^+\pi^0$, $\overline B^0_s\to D^0\pi^0$, $\overline B^0_s\to D^+\pi^-$, $\overline B^0_s\to \pi^+\pi^-$ and $\overline B^0_s\to \pi^0\pi^0$ modes to test the isospin symmetry.
The branching fraction of $\overline B^0_s\to \pi^+\pi^-$ mode has been measured by many experiments and averaged to be $(7.0\pm 1.0)\times 10^{-7}$ \cite{PDG}.
And the upper limits of $\mathcal{B}r(D^+_s\to \pi^+\pi^0)$ and $\mathcal{B}r(\overline B^0_s\to \pi^0\pi^0)$ are given by $1.2\times 10^{-4}$ and $2.1\times 10^{-4}$, respectively \cite{PDG}.
It is significant to perform a more precise measurement for above five channels in the future.

\section{Summary}\label{summary}

Flavor symmetry is a model-independent tool to analyze heavy meson and baryon decays.
The flavor invariants are independent of the detailed dynamics and determined by fitting experimental data.
In this work, we propose a simple algorithm to generate the isospin, $V$-spin and $U$-spin sum rules of heavy hadron decays.
We found that the effective Hamiltonian of heavy quark decay is fully invariant under a series of lowering operators $I_-^n$, $V_-^n$ and $U_-^n$.
The isospin, $V$-spin and $U$-spin sum rules can be generated from several master formulas without the Wigner-Eckhart invariants.
Taking the two-body decays of $D$ and $B$ mesons as examples, our approach is presented in detail.
In addition, we suggest to measure the branching fractions of $D^+_s\to \pi^+\pi^0$, $\overline B^0_s\to D^0\pi^0$, $\overline B^0_s\to D^+\pi^-$, $\overline B^0_s\to \pi^+\pi^-$ and $\overline B^0_s\to \pi^0\pi^0$ modes to test the isospin symmetry.

\begin{acknowledgements}

This work was supported in part by the National Natural Science Foundation of China under Grants No. 12105099.

\end{acknowledgements}

\appendix

\section{$V$-spin sum rules}\label{Vsum}

In this appendix, we derive the $V$-spin sum rules in the $D\to PP$, $\overline B\to DP$ and $\overline B\to PP$ modes.
The $V$-spin lowering operator $V_-$ is
\begin{eqnarray}
 V_-= \left( \begin{array}{ccc}
   0   & 0  & 0 \\
     0 &  0  & 0 \\
    1 & 0 & 0 \\
  \end{array}\right).
\end{eqnarray}
In the charm quark decay, $[H^{(0)}(\overline 6)]$, $[H^{(0)}(15)]_i$, $[H^{(0,1)}(3_{t,p})]$  are transformed under $V_-^n$ as
\begin{align}
 V_-[H^{(0)}(\overline 6)] =0, \qquad  V_-[H^{(0)}(15)]_{2,3} =0, \qquad
V_-[H^{(0,1)}( 3_{t,p})] = 0,
\end{align}
\begin{align}
 V_-^2[H^{(0)}(15)]_1 =V_-   \left( \begin{array}{ccc}
   0   & 0  & 0 \\
     8V_{cs}^*V_{ud} & 0 & 0 \\
    8V_{cs}^*V_{us} & 0 & 0 \\
  \end{array}\right)=0.
\end{align}
So the $V$-spin sum rules of DCS transition can be generated by $V_-^n$ if $n\geq 1$, and the $V$-spin sum rules of CF and SCS transitions can be generated by $V_-^n$ if $n\geq 2$.
The coefficient matrix $[V_-]_D$ is derived to be
\begin{eqnarray}
 [V_-]_D= \left( \begin{array}{ccc}
   0   & 0  & 0 \\
     0 &  0  & 0 \\
    1 & 0 & 0 \\
  \end{array}\right).
\end{eqnarray}
The coefficient matrix $[V_-]_{P_8}$ is derived to be
\begin{eqnarray}
 [V_-]_{P_8}= \left( \begin{array}{cccccccc}
  0 & 0& 0& 0& 0& 0& 0& 0 \\
  0& 0& 0& -\frac{1}{\sqrt{2}}& 0& 0& 0& 0 \\
 0& 0& 0& 0& -1& 0& 0& 0 \\
  0& 0& 0& 0& 0& 0& 0& 0 \\
  0& 0& 0& 0& 0& 0& 0& 0\\
 1& 0& 0& 0& 0& 0& 0& 0\\
 0& \frac{1}{\sqrt{2}}& 0& 0& 0& 0& 0& \frac{\sqrt{6}}{2} \\
 0& 0& 0& -\frac{\sqrt{6}}{2}& 0&0& 0& 0 \\
  \end{array}\right).
\end{eqnarray}
The sum of decay amplitudes generated from $D_\gamma\to P_\alpha P_\beta$ under $V_-$ is
\begin{align}\label{rule4}
{ SumV_-}\,[\gamma, \alpha,\beta]= \sum_\mu\left[\{[V_-]_{P_8}\}_\alpha^\mu \mathcal{A}_{ \gamma \to \mu \beta} +  \{[V_-]_{P_8}\}_\beta^\mu \mathcal{A}_{\gamma\to \alpha\mu } + \{[V_-]_{D}\}_\gamma^\mu \mathcal{A}_{\mu\to \alpha \beta }\right].
\end{align}
The $V$-spin sum rules in the $D\to PP$ modes are derived to be
\begin{align}
{ SumV_-^2}\,[D^0, \pi^+,K^+]&=-\frac{{ SumV_-}\,[D^0, \pi^+,\pi^0]}{\sqrt{2}}-\sqrt{\frac{3}{2}}\,{ SumV_-}\,[D^0, \pi^+,\eta_8]\nonumber\\&~~~~~+{ SumV_-}\,[D^0, \overline K^0,K^+]+{ SumV_-}\,[D^+_s, \pi^+,K^+]\nonumber\\&=2\, \mathcal{A}(D^+_s\to K^+ \overline K^0)-\sqrt{6}\,\mathcal{A}(D^+_s\to \pi^+\eta_8)-\sqrt{2}\,\mathcal{A}(D^+_s\to \pi^+ \pi^0)\nonumber\\&~~~~~-\sqrt{6}\, \mathcal{A}(D^0\to \overline K^0 \eta_8)-\sqrt{2}\,\mathcal{A}(D^0\to \pi^0\overline K^0)\nonumber\\&~~~~~~~-2\,\mathcal{A}(D^0\to \pi^+ K^-)=0,
\end{align}
\begin{align}
{ SumV_-^2}\,[D^0, K^+,K^+]&=-\frac{{ SumV_-}\,[D^0, \pi^0,K^+]}{\sqrt{2}}-\frac{{ SumV_-}\,[D^0, K^+,\pi^0]}{\sqrt{2}}-\sqrt{\frac{3}{2}}\,{ SumV_-}\,[D^0, K^+,\eta_8]\nonumber\\&~~~~~-\sqrt{\frac{3}{2}}\,{ SumV_-}\,[D^0,\eta_8, K^+]+{ SumV_-}\,[D^+_s, K^+,K^+]\nonumber\\&=-2\sqrt{6}\,\mathcal{A}(D^+_s\to K^+ \eta_8)-2\sqrt{2}\,\mathcal{A}(D^+_s\to \pi^0K^+)+3\,\mathcal{A}(D^0\to \eta_8 \eta_8)\nonumber\\&~~~~~-4\, \mathcal{A}(D^0\to K^+ K^-)+2\sqrt{3}\,\mathcal{A}(D^0\to \pi^0\eta_8)\nonumber\\&~~~~~~~+\mathcal{A}(D^0\to \pi^0\pi^0 )=0,
\end{align}
\begin{align}
{ SumV_-}\,[D^0, K^+, K^0]&=\mathcal{A}(D^+_s\to K^+ K^0)-\sqrt{\frac{3}{2}}\,\mathcal{A}(D^0\to K^0\eta_8)\nonumber\\&~~~~~-\frac{\mathcal{A}(D^0\to \pi^0 K^0)}{\sqrt{2}}-\mathcal{A}(D^0\to \pi^-K^+)=0.
\end{align}

In the $b\to c\bar u q$ transition, $[H^{(0)}(8)]$ is transformed under $V_-^n$ as
\begin{align}
 V_-^3[ H^{(0)}(8)] & =V_- \left\{V_-  \left( \begin{array}{ccc}
   -V_{cb}V_{us}^*   & 0  & 0 \\
     0 & 0 & 0 \\
    0 & V_{cb}V_{ud}^* & V_{cb}V_{us}^* \\
  \end{array}\right)\right\}=V_-  \left( \begin{array}{ccc}
   0  & 0  & 0 \\
     0 & 0 & 0 \\
    -2V_{cb}V_{us}^* & 0 & 0 \\
  \end{array}\right)= 0.
\end{align}
So the $V$-spin sum rules of $b\to c\bar u d$ transition can be generated by $V_-^n$ if $n\geq 2$, and the $V$-spin sum rules of $b\to c\bar u s$ transition can be generated by $V_-^n$ if $n\geq 3$.
Under the $V$-spin lowering operator $V_-$, we have $[V_-]_{\overline B} = [V_-]_{D}$.
The sum of decay amplitudes generated from $\overline B_\gamma\to D_\alpha P_\beta$ under $V_-$ is
\begin{align}\label{rule5}
 { SumV_-}\,[\gamma, \alpha,\beta]=  \sum_\mu\left[\{[V_-]^T_{D}\}_\alpha^\mu \mathcal{A}_{ \gamma \to \mu \beta} +  \{[V_-]_{P_8}\}_\beta^\mu \mathcal{A}_{\gamma\to \alpha\mu } + \{[V_-]_{\overline B}\}_\gamma^\mu \mathcal{A}_{\mu\to \alpha \beta }\right].
\end{align}
The $V$-spin sum rules in the $\overline B\to DP$ modes are derived to be
\begin{align}
{ SumV_-^2}\,[B^-, D^+_s, K^0]&=-{ SumV_-}\,[B^-, D^0,K^0]-{ SumV_-}\,[B^-, D^+_s,\pi^-]+{ SumV_-}\,[\overline B^0_s, D^+_s, K^0]\nonumber\\&=-2\,\big[ \mathcal{A}(\overline B^0_s\to D^+_s\pi^-)+\,\mathcal{A}(\overline B^0_s\to D^0 K^0)-\mathcal{A}( B^-\to D^0\pi^-)\big]=0,
\end{align}
\begin{align}
{ SumV_-^2}\,[\overline B^0, D^+_s, K^+]&=-{ SumV_-}\,[\overline B^0, D^0,K^+]-\frac{{ SumV_-}\,[\overline B^0, D^+_s,\pi^0]}{\sqrt{2}}\nonumber\\&~~~~~-\sqrt{\frac{3}{2}}\,{ SumV_-}\,[\overline B^0, D^+_s, \eta_8]\nonumber\\&=-2\, \mathcal{A}(\overline B^0\to D^+_sK^-)+\sqrt{6}\,\mathcal{A}(\overline B^0\to D^0 \eta_8)\nonumber\\&~~~~~+\sqrt{2}\,\mathcal{A}( \overline B^0\to D^0\pi^0)=0,
\end{align}
\begin{align}
{ SumV_-^3}\,[B^-, D^+_s,K^+]&=-{ SumV_-^2}\,[B^-, D^0,K^+]-\frac{{ SumV_-^2}\,[B^-, D^+_s,\pi^0]}{\sqrt{2}}\nonumber\\&~~~~~-\sqrt{\frac{3}{2}}\,{ SumV_-^2}\,[B^-, D^+_s,\eta_8]+{ SumV_-^2}\,[\overline B^0_s, D^+_s,K^+]\nonumber\\&=\sqrt{2}\,{ SumV_-}\,[B^-, D^0,\pi^0]+\sqrt{6}\,{ SumV_-}\,[B^-, D^0,\eta_8]\nonumber\\&~~~~~-2\,{ SumV_-}\,[B^-, D^+_s,K^-]-2\,{ SumV_-}\,[\overline B^0_s, D^0,K^+]\nonumber\\&~~~~~-\sqrt{2}\,{ SumV_-}\,[\overline B^0_s, D^+_s,\pi^0]-\sqrt{6}\,{ SumV_-}\,[\overline B^0_s, D^+_s,\eta_8]\nonumber\\&=3\,\big[-2\mathcal{A}(\overline B^0_s\to D^+_sK^-)+\sqrt{6}\,\mathcal{A}(\overline B^0_s\to D^0\eta_8)\nonumber\\&~~~~~+\sqrt{2}\,\mathcal{A}(\overline B^0_s\to D^0\pi^0)+2\,\mathcal{A}(B^-\to D^0 K^-)\big]=0.
\end{align}

In the $b\to u\bar u q$ transition, $[H^{(0,u)}(\overline 6)]$, $[H^{(0,u)}(15)]_i$, $[H(3)]$ are transformed under $V_-^n$ as
\begin{align}
 V_-^2[H^{(0,u)}(\overline 6)] =V_-   \left( \begin{array}{ccc}
   0   & 0  & 0 \\
     0 & 0 & 0 \\
    0 & -2V_{ub}V_{us}^* & 2V_{ub}V_{ud}^* \\
  \end{array}\right)=0,
\end{align}
\begin{align}
 V_-^3[H^{(0,u)}(15)]_1 & =V_- \left\{V_-  \left( \begin{array}{ccc}
   6V_{ub}V_{us}^*   & 0  & 0 \\
     0 & -V_{ub}V_{us}^* & 0 \\
     0 & -4V_{ub}V_{ud}^* & -5V_{ub}V_{us}^* \\
  \end{array}\right)\right\}=V_-  \left( \begin{array}{ccc}
   0  & 0  & 0 \\
       0 & 0 & 0 \\
     -16V_{ub}V_{us}^* & 0 & 0 \\
  \end{array}\right)= 0,
\end{align}
\begin{align}
 V_-^2[H^{(0,u)}(15)]_2 & =  V_-\{V_-[H^{(0,u)}(15)]_2\} =V_-  \left( \begin{array}{ccc}
   0   & 0  & 0 \\
     -V_{ub}V_{us}^* & 0 & 0 \\
    -4V_{ub}V_{ud}^* & 0 & 0 \\
  \end{array}\right) = 0,
\end{align}
\begin{align}
 V_-^2[H^{(0,u)}(15)]_3 & =  V_-  \left( \begin{array}{ccc}
   0   & 0  & 0 \\
     0 & 0 & 0 \\
    -5V_{ub}V_{us}^* & 0 & 0 \\
  \end{array}\right) = 0,
\end{align}
\begin{align}
 V_-^2[H(3)] =V_-  ( \begin{array}{ccc}
   b   & 0  & 0
  \end{array}) = 0.
\end{align}
So the $V$-spin sum rules of $b\to u\overline u d$ ($b\to u\overline u s$) transition can be generated by $V_-^n$ if $n\geq 2$ ($n\geq 3$).
The sum of decay amplitudes generated from $\overline B_\gamma\to P_\alpha P_\beta$ under $V_-$ is
\begin{align}\label{rule6}
 { SumV_-}\,[\gamma, \alpha,\beta]=  \sum_\mu\left[\{[V_-]_{P_8}\}_\alpha^\mu \mathcal{A}_{ \gamma \to \mu \beta} +  \{[V_-]_{P_8}\}_\beta^\mu \mathcal{A}_{\gamma\to \alpha\mu } + \{[V_-]_{\overline B}\}_\gamma^\mu \mathcal{A}_{\mu\to \alpha \beta }\right].
\end{align}
With Eq.~\eqref{rule6}, the $V$-spin sum rules in the $\overline B\to PP$ modes are derived to be
\begin{align}
{ SumV_-^2}\,[B^-, K^+, K^0]&=-\frac{{ SumV_-}\,[B^-, \pi^0, K^0]}{\sqrt{2}}-{ SumV_-}\,[ B^-, K^+,\pi^-]\nonumber\\&~~~~~-\sqrt{\frac{3}{2}}\,{ SumV_-}\,[B^-, \eta_8,K^0]+{ SumV_-}\,[\overline B^0_s, K^+,K^0]\nonumber\\&=-\sqrt{6}\, \mathcal{A}(\overline B^0_s\to K^0\eta_8)-\sqrt{2}\,\mathcal{A}(\overline B^0_s\to \pi^0 K^0)\nonumber\\&~~~~-2\,\mathcal{A}(\overline B^0_s\to \pi^- K^+)-2\,\mathcal{A}( B^-\to K^0K^-)\nonumber\\&~~~~~~+\sqrt{6}\,\mathcal{A}( B^-\to \pi^- \eta_8)+\sqrt{2}\,\mathcal{A}( B^-\to \pi^0\pi^-)=0,
\end{align}
\begin{align}
{ SumV_-^2}\,[\overline B^0, K^+, K^+]&=-\frac{{ SumV_-}\,[\overline B^0, \pi^0, K^+]}{\sqrt{2}}-\frac{{ SumV_-}\,[ \overline B^0, K^+, \pi^0]}{\sqrt{2}}\nonumber\\&~~~~~-\sqrt{\frac{3}{2}}\,{ SumV_-}\,[\overline B^0, \eta_8,K^+]-\sqrt{\frac{3}{2}}\,{ SumV_-}\,[\overline B^0,K^+, \eta_8]\nonumber\\&=3\, \mathcal{A}(\overline B^0\to \eta_8\eta_8)-4\,\mathcal{A}(\overline B^0\to K^+K^- )\nonumber\\&~~~~+2\sqrt{3}\,\mathcal{A}(\overline B^0\to \pi^0 \eta_8)+\mathcal{A}( \overline B^0\to \pi^0\pi^0)=0,
\end{align}
\begin{align}
{ SumV_-^3}\,[B^-, K^+,K^+]&={ SumV_-^2}\,[ \overline B^0_s,K^+, K^+]-\frac{{ SumV_-^2}\,[B^-, \pi^0,K^+]}{\sqrt{2}}-\frac{{ SumV_-^2}\,[B^-,K^+, \pi^0]}{\sqrt{2}}\nonumber\\&~~~-\sqrt{\frac{3}{2}}\,{ SumV_-^2}\,[ B^-, K^+,\eta_8]-\sqrt{\frac{3}{2}}\,{ SumV_-^2}\,[ B^-,\eta_8, K^+]\nonumber\\&={ SumV_-}\,[B^-, \pi^0,\pi^0]+\sqrt{3}\,{ SumV_-}\,[B^-, \pi^0,\eta_8]-2\,{ SumV_-}\,[B^-, K^+,K^-]\nonumber\\&~~~~~-2\,{ SumV_-}\,[B^-, K^-,K^+]+\sqrt{3}\,{ SumV_-}\,[B^-, \eta_8,\pi^0]\nonumber\\&~~~~~~+3\,{ SumV_-}\,[B^-, \eta_8,\eta_8]-\sqrt{2}\,{ SumV_-}\,[\overline B^0_s, \pi^0,K^+]\nonumber\\&~~~~~~~-\sqrt{2}\,{ SumV_-}\,[\overline B^0_s,K^+, \pi^0]-\sqrt{6}\,{ SumV_-}\,[\overline B^0_s,K^+, \eta_8]\nonumber\\&~~~~~~~~-\sqrt{6}\,{ SumV_-}\,[\overline B^0_s, \eta_8,K^+]\nonumber\\&=3\,\big[\,3\, \mathcal{A}(\overline B^0_s\to \eta_8\eta_8)-4\,\mathcal{A}(\overline B^0_s\to K^+K^-)+2\sqrt{3}\,\mathcal{A}(\overline B^0_s\to \pi^0\eta_8)\nonumber\\&~~~~~+\mathcal{A}(\overline B^0_s\to \pi^0\pi^0)+2\sqrt{6}\,\mathcal{A}( B^-\to K^-\eta_8)\nonumber\\&~~~~~~~+2\sqrt{2}\,\mathcal{A}( B^-\to K^-\pi^0)\big]=0.
\end{align}

\section{$U$-spin sum rules}\label{Usum}
In this appendix, we derive the $U$-spin sum rules in the $D\to PP$, $\overline B\to DP$ and $\overline B\to PP$ modes.
The $U$-spin lowering operator $U_-$ is
\begin{eqnarray}
 U_-= \left( \begin{array}{ccc}
   0   & 0  & 0 \\
     0 &  0  & 0 \\
    0 & 1 & 0 \\
  \end{array}\right).
\end{eqnarray}
In charm quark decay, $[H^{(0)}(\overline 6)]$, $[H^{(0)}(15)]_i$, $[H^{(0,1)}(3_{t,p})]$  are transformed under $U_-^n$ as
\begin{align}
 U^2_-[H^{(0)}(\overline 6)] =U_- \left( \begin{array}{ccc}
   0   & 0  & 0 \\
     0 & 0 & 0 \\
    0 & -4V_{cs}^*V_{ud} & 2(V_{cd}^*V_{ud}-V_{cs}^*V_{us}) \\
  \end{array}\right)=0,
\end{align}
\begin{align}
U_-[H^{(0,1)}( 3_{t,p})] = 0,
\end{align}
\begin{align}
 U_-^3[H^{(0)}(15)]_1 & =U_- \left\{U_-  \left( \begin{array}{ccc}
   0   & 0  & 0 \\
     0 & 4V_{cs}^*V_{ud} & 0 \\
     0 & 4(V_{cs}^*V_{us}-V_{cd}^*V_{ud}) & -4V_{cs}^*V_{ud} \\
  \end{array}\right)\right\} \nonumber\\&~~=U_-  \left( \begin{array}{ccc}
   0  & 0  & 0 \\
       0 & 0 & 0 \\
     -4V_{cs}^*V_{ud} & -4V_{cs}^*V_{ud} & 0 \\
  \end{array}\right)= 0,
\end{align}

\begin{align}
 U_-^3[H^{(0)}(15)]_2 & =U_- \left\{U_-  \left( \begin{array}{ccc}
   0   & 0  & 0 \\
     4V_{cs}^*V_{ud} & 0 & 0 \\
     4(V_{cs}^*V_{us}-V_{cd}^*V_{ud}) & 0 & 0 \\
  \end{array}\right)\right\}=U_-  \left( \begin{array}{ccc}
   0  & 0  & 0 \\
       0 & 0 & 0 \\
     -8V_{cs}^*V_{ud} & 0 & 0 \\
  \end{array}\right)= 0,
\end{align}

\begin{align}
 U_-^2[H^{(0)}(15)]_3 & =U_-  \left( \begin{array}{ccc}
   0   & 0  & 0 \\
     0 & 0 & 0 \\
     -4V_{cs}^*V_{ud} & 0 & 0 \\
  \end{array}\right) = 0,
\end{align}
The $U$-spin sum rules of DCS, SCS and CF transitions are generated by $U_-^n$ if $n\geq 1$, $n\geq 2$ and $n\geq 3$, respectively.
The coefficient matrix $[U_-]_D$ is derived to be
\begin{eqnarray}
 [U_-]_D= \left( \begin{array}{ccc}
   0   & 0  & 0 \\
     0 &  0  & 0 \\
    0 & 1 & 0 \\
  \end{array}\right).
\end{eqnarray}
The coefficient matrix $[U_-]_{P_8}$ is derived to be
\begin{eqnarray}
 [U_-]_{P_8}= \left( \begin{array}{cccccccc}
  0 & 0& 0& -1& 0& 0& 0& 0 \\
  0& 0& 0& 0& \frac{1}{\sqrt{2}}& 0& 0& 0 \\
 0& 0& 0& 0& 0& 0& 0& 0 \\
  0& 0& 0& 0& 0& 0& 0& 0 \\
  0& 0& 0& 0& 0& 0& 0& 0\\
 0& -\frac{1}{\sqrt{2}}& 0& 0& 0& 0& 0& \frac{\sqrt{6}}{2}\\
 0& 0& 1& 0& 0& 0& 0& 0 \\
 0& 0& 0&0 & -\frac{\sqrt{6}}{2}&0& 0& 0 \\
  \end{array}\right).
\end{eqnarray}
The sum of decay amplitudes generated from $D_\gamma\to P_\alpha P_\beta$ under $U_-$ is
\begin{align}\label{rule7}
{ SumU_-}\,[\gamma, \alpha,\beta]= \sum_\mu\left[\{[U_-]_{P_8}\}_\alpha^\mu \mathcal{A}_{ \gamma \to \mu \beta} +  \{[U_-]_{P_8}\}_\beta^\mu \mathcal{A}_{\gamma\to \alpha\mu } + \{[U_-]_{D}\}_\gamma^\mu \mathcal{A}_{\mu\to \alpha \beta }\right].
\end{align}
The $U$-spin sum rules in the $D\to PP$ modes are derived to be
\begin{align}
{ SumU^3_-}\,[D^+, K^+, K^0]&={ SumU^2_-}\,[D^+_s\to K^+ K^0]-\sqrt{\frac{3}{2}}\,{ SumU^2_-}\,[D^+\to K^+\eta_8]\nonumber\\&~~~~~+\frac{{ SumU^2_-}\,[D^+\to \pi^0 K^+]}{\sqrt{2}}-{ SumU^2_-}\,[D^+\to \pi^+K^0]\nonumber\\&= -\sqrt{6}\,{ SumU_-}\,[D^+_s\to K^+\eta_8]+\sqrt{2}\,{ SumU_-}\,[D^+_s\to \pi^0K^+]\nonumber\\&~~~~~-2\,{ SumU_-}\,[D^+_s\to \pi^+K^0]-2\,{ SumU_-}\,[D^+\to K^+\overline K^0]\nonumber\\&~~~~~~~+\sqrt{6}\,{ SumU_-}\,[D^+\to \pi^+\eta_8]-\sqrt{2}\,{ SumU_-}\,[D^+\to \pi^+\pi^0]\nonumber\\&=-6\,\mathcal{A}(D^+_s\to K^+\overline K^0)+3\sqrt{6}\,\mathcal{A}(D^+_s\to \pi^+\eta_8)\nonumber\\&~~~~~ -3\sqrt{2}\,\mathcal{A}(D^+_s\to \pi^+\pi^0)+6\,\mathcal{A}(D^+\to \pi^+\overline K^0)
=0,
\end{align}
\begin{align}
{ SumU^3_-}\,[D^0, K^0, K^0]&=-\sqrt{6}\,{ SumU^2_-}\,[D^0\to K^0 \eta_8]+\sqrt{2}\,{ SumU^2_-}\,[D^0\to \pi^0K^0]\nonumber\\&=3\,{ SumU_-}\,[D^0\to \eta_8\eta_8]-4\,{ SumU_-}\,[D^0\to K^0\overline K^0]\nonumber\\&~~~~~-2\sqrt{3}\,{ SumU_-}\,[D^0\to \pi^0\eta_8]+{ SumU_-}\,[D^0\to \pi^0\pi^0]=\nonumber\\&=6\sqrt{6}\,\mathcal{A}(D^0\to \overline K^0\eta_8)-6\sqrt{2}\,\mathcal{A}(D^0\to \pi^0\overline K^0)=0,
\end{align}
\begin{align}
{ SumU^2_-}\,[D^+, K^+, K^0]&={ SumU_-}\,[D^+_s\to K^+ K^0]-\sqrt{\frac{3}{2}}\,{ SumU_-}\,[D^+\to K^+\eta_8]\nonumber\\&~~~~~+\frac{{ SumU_-}\,[D^+\to \pi^0 K^+]}{\sqrt{2}}-{ SumU_-}\,[D^+\to \pi^+K^0]\nonumber\\&= -\sqrt{6}\,\mathcal{A}(D^+_s\to K^+\eta_8)+\sqrt{2}\,\mathcal{A}(D^+_s\to \pi^0K^+)-2\,\mathcal{A}(D^+_s\to \pi^+K^0)\nonumber\\&~~~~~-2\,\mathcal{A}(D^+\to K^+\overline K^0)+\sqrt{6}\,\mathcal{A}(D^+\to \pi^+\eta_8)\nonumber\\&~~~~~~~-\sqrt{2}\,\mathcal{A}(D^+\to \pi^+\pi^0)=
0,
\end{align}
\begin{align}
{ SumU^2_-}\,[D^0, K^0, K^0]&=-\sqrt{6}\,{ SumU_-}\,[D^0\to K^0 \eta_8]+\sqrt{2}\,{ SumU_-}\,[D^0\to \pi^0K^0]\nonumber\\&=3\,\mathcal{A}(D^0\to \eta_8\eta_8)-4\,\mathcal{A}(D^0\to K^0\overline K^0)\nonumber\\&~~~~~-2\sqrt{3}\,\mathcal{A}(D^0\to \pi^0\eta_8)+\mathcal{A}(D^0\to \pi^0\pi^0)=0,
\end{align}
\begin{align}
{ SumU_-}\,[D^+, K^+, K^0]&=\mathcal{A}(D^+_s\to K^+ K^0)-\sqrt{\frac{3}{2}}\,\mathcal{A}(D^+\to K^+\eta_8)\nonumber\\&~~~~~+\frac{\mathcal{A}(D^+\to \pi^0 K^+)}{\sqrt{2}}-\mathcal{A}(D^+\to \pi^+K^0)=0,
\end{align}
\begin{align}
{ SumU_-}\,[D^0, K^0, K^0]&=-\sqrt{6}\,\mathcal{A}(D^0\to K^0 \eta_8)+\sqrt{2}\,\mathcal{A}(D^0\to \pi^0K^0)=0.
\end{align}
One should note the $U$-spin sum rules derived by $U^n_-$ do not dependent on the  Wolfenstein approximation of the CKM matrix.

In the $b\to c\bar u q$ transition, $[H^{(0)}(8)]$ is transformed under $U_-^n$ as
\begin{align}
 U_-^2[ H^{(0)}(8)] & =U_-  \left( \begin{array}{ccc}
   0  & -V_{cb}V_{us}^*  & 0 \\
     0 & 0 & 0 \\
    0 & 0 & 0 \\
  \end{array}\right)= 0.
\end{align}
So the $U$-spin sum rules of $b\to c\bar u d$ transition are generated by $U_-^n$ if $n\geq 1$, and the $U$-spin sum rules of $b\to c\bar u s$ transition are generated by $U_-^n$ if $n\geq 2$.
Under the $U$-spin lowering operator $U_-$, $[U_-]_{\overline B} = [U_-]_{D}$.
The sum of decay amplitudes generated from $\overline B_\gamma\to D_\alpha P_\beta$ under $U_-$ is
\begin{align}\label{rule8}
 { SumU_-}\,[\gamma, \alpha,\beta]=  \sum_\mu\left[\{[U_-]^T_{D}\}_\alpha^\mu \mathcal{A}_{ \gamma \to \mu \beta} +  \{[U_-]_{P_8}\}_\beta^\mu \mathcal{A}_{\gamma\to \alpha\mu } + \{[U_-]_{\overline B}\}_\gamma^\mu \mathcal{A}_{\mu\to \alpha \beta }\right].
\end{align}
The $U$-spin sum rules in the $\overline B\to DP$ modes are
\begin{align}
{ SumU_-}\,[B^0, D^0, K^0]&= \mathcal{A}(\overline B^0_s\to D^0K^0)-\sqrt{\frac{3}{2}}\,\mathcal{A}(\overline B^0\to D^0 \eta_8)+\frac{\mathcal{A}(\overline B^0\to D^0 \pi^0)}{\sqrt{2}}=0,
\end{align}
\begin{align}
{ SumU_-}\,[B^0, D^+_s, \pi^-]&= \mathcal{A}(\overline B^0_s\to D^+_s \pi^-)+\mathcal{A}(\overline B^0\to D^+_s K^-)-\mathcal{A}(\overline B^0\to D^+ \pi^-)=0,
\end{align}
\begin{align}
{ SumU^2_-}\,[B^0, D^0, K^0]&= { SumU_-}\,[\overline B^0_s\to D^0K^0]-\sqrt{\frac{3}{2}}\,{ SumU_-}\,[\overline B^0\to D^0 \eta_8]\nonumber\\&~~~~~+\frac{{ SumU_-}\,[\overline B^0\to D^0 \pi^0]}{\sqrt{2}}\nonumber\\&=\sqrt{2}\,\mathcal{A}(\overline B^0_s\to D^0\pi^0)-\sqrt{6}\,\mathcal{A}(\overline B^0_s\to D^0\eta_8)-2\,\mathcal{A}(\overline B^0\to D^0 \overline K^0)=0,
\end{align}
\begin{align}
{ SumU^2_-}\,[B^0, D^+_s, \pi^-]&= { SumU_-}\,[\overline B^0_s\to D^+_s \pi^-]+{ SumU_-}\,[\overline B^0\to D^+_s K^-]\nonumber\\&~~~~~-{ SumU_-}\,[\overline B^0\to D^+ \pi^-]\nonumber\\&=2\,\Big[\mathcal{A}(\overline B^0_s\to D^+_s K^-)-\mathcal{A}(\overline B^0_s\to D^+ \pi^-)-\mathcal{A}(\overline B^0\to D^+ K^-)\Big]
=0.
\end{align}

In the $b\to u\bar u q$ transition, $[H^{(0,u)}(\overline 6)]$, $[H^{(0,u)}(15)]_i$, $[H(3)]$ are transformed under $U_-^n$ as
\begin{align}
 U_-^2[H^{(0,u)}(\overline 6)] =U_-   \left( \begin{array}{ccc}
   0   & 0  & 0 \\
     0 & 0 & 0 \\
    -2V_{ub}V_{us}^* & 0 & 0 \\
  \end{array}\right)=0,
\end{align}
\begin{align}
 U_-^2[H^{(0,u)}(15)]_1 & =U_-  \left( \begin{array}{ccc}
   0  & 3V_{ub}V_{us}^*  & 0 \\
       0 & 0 & 0 \\
     0 & 0 & 0 \\
  \end{array}\right)= 0,
\end{align}
\begin{align}
 U_-^2[H^{(0,u)}(15)]_2 & =U_-  \left( \begin{array}{ccc}
   3V_{ub}V_{us}^*   & 0  & 0 \\
     0 & -2V_{ub}V_{us}^* & 0 \\
    0 & 0 & -V_{ub}V_{us}^* \\
  \end{array}\right) = 0,
\end{align}
\begin{align}
 U_-^2[H^{(0,u)}(15)]_3 & =U_-  \left( \begin{array}{ccc}
   0   & 0  & 0 \\
     0 & 0 & 0 \\
    0 & -V_{ub}V_{us}^* & 0 \\
  \end{array}\right) = 0,
\end{align}
\begin{align}
 U_-^2[H(3)] =U_-  ( \begin{array}{ccc}
   0   & b  & 0
  \end{array}) = 0.
\end{align}
So the $U$-spin sum rules of $b\to u\overline u d$ ($b\to u\overline u s$) transition can be generated by $U_-^n$ if $n\geq 1$ ($n\geq 2$).
The sum of decay amplitudes generated from $\overline B_\gamma\to P_\alpha P_\beta$ under $U_-$ is
\begin{align}\label{rule9}
 { SumU_-}\,[\gamma, \alpha,\beta]=  \sum_\mu\left[\{[U_-]_{P_8}\}_\alpha^\mu \mathcal{A}_{ \gamma \to \mu \beta} +  \{[U_-]_{P_8}\}_\beta^\mu \mathcal{A}_{\gamma\to \alpha\mu } + \{[U_-]_{\overline B}\}_\gamma^\mu \mathcal{A}_{\mu\to \alpha \beta }\right].
\end{align}
With Eq.~\eqref{rule9}, the $U$-spin sum rules of $\overline B\to PP$ modes are derived to be
\begin{align}
{ SumU_-}\,[B^-, \pi^-, K^0]&= \mathcal{A}( B^-\to K^0K^-)-\sqrt{\frac{3}{2}}\,\mathcal{A}(B^-\to \pi^- \eta_8)+\frac{\mathcal{A}( B^-\to \pi^0\pi^-)}{\sqrt{2}}=0,
\end{align}
\begin{align}
{ SumU_-}\,[B^0, K^+, \pi^-]&= \mathcal{A}(\overline B^0_s\to  \pi^-K^+)+\mathcal{A}(\overline B^0\to K^+ K^-)-\mathcal{A}(\overline B^0\to \pi^+ \pi^-)=0,
\end{align}
\begin{align}
{ SumU^2_-}\,[B^-, \pi^-, K^0]&= { SumU_-}\,[B^-\to K^0K^-]-\sqrt{\frac{3}{2}}\,{ SumU_-}\,[B^-\to \pi^- \eta_8]\nonumber\\&~~~~~+\frac{{ SumU_-}\,[B^-\to \pi^0\pi^-]}{\sqrt{2}}\nonumber\\&=\sqrt{2}\,\mathcal{A}(B^-\to \pi^0K^-)-\sqrt{6}\,\mathcal{A}(B^-\to K^-\eta_8)\nonumber\\&~~~~~-2\,\mathcal{A}(B^-\to \pi^- \overline K^0)=0,
\end{align}
\begin{align}
{ SumU^2_-}\,[B^0, K^+, \pi^-]&= { SumU_-}\,[\overline B^0_s\to  \pi^-K^+]+{ SumU_-}\,[\overline B^0\to K^+ K^-]\nonumber\\&~~~~~-{ SumU_-}\,[\overline B^0\to \pi^+ \pi^-]\nonumber\\&=2\,\Big[\mathcal{A}(\overline B^0_s\to K^+ K^-)-\mathcal{A}(\overline B^0_s\to \pi^+ \pi^-)-\mathcal{A}(\overline B^0\to \pi^+ K^-)\Big]
=0.
\end{align}



\end{document}